\documentclass{aastex63}
\usepackage{amsmath}
\usepackage{amssymb}

\received{XX}
\revised{YY}
\accepted{ZZ}

\submitjournal{ApJ}

\shorttitle{Chemistry of ro-vibrationally excited H$_2$ in disks}
\shortauthors{Ruaud et al.}

\begin{document}

\title{H$_2$ ro-vibrational excitation in protoplanetary disks and its effects on the chemistry}

\correspondingauthor{Maxime Ruaud}
\email{maxime.ruaud@gmail.com}

\author[0000-0002-0786-7307]{Maxime Ruaud}
\affiliation{NASA Ames Research Center, Moffett Blvd., Mountain View, CA 94035, USA}
\affiliation{Carl Sagan Center, SETI Institute, Mountain View, CA 94035, USA}

\begin{abstract}

The effect of H$_2$ ro-vibrational excitation on the chemistry of protoplanetary disks is studied using a framework that solves for the disk physical and chemical structure and includes a detailed calculation of H$_2$ level populations. Chemistry with ro-vibrationally excited H$_2$ is found to be important for the formation of several commonly observed species in disks and this work demonstrates the need to accurately treat PDR chemistry in disks if we are to make inferences on the chemical state of the disk during planet formation epochs. This is found to be even more critical for molecules like C$_2$H, CN or HCN that are commonly used to infer changes in the elemental disk C/O and N/O ratios, with implications for planetesimal formation and the composition of exoplanet atmospheres. 
Computed vertical column densities with the full H$_2$ population calculation are increased by $\sim1-2$ orders of magnitude for molecules such as CN, HCN/HNC compared to calculations with no treatment of excited H$_2$. For the commonly used pseudo-level approximation, the computed columns of these molecules are overestimated by a factor of $\sim3-5$ when compared to the full model. We further note that the computed abundance for these molecules strongly depends on the strength of the FUV photons at energies that pump H$_2$ (i.e. 11-13.6 eV), which is not well constrained in disks, and that rate constants as a function of H$_2$ ro-vibrational levels for the key reaction N + H$_2\rightarrow $ NH are needed for a more accurate assessment of CN/HCN chemistry but are currently unavailable. 

\end{abstract}

\keywords{Protoplanetary disks (1300), Astrochemistry (75), Chemical abundances (224),  Star formation (1569)}

\section{Introduction}
\label{sec:intro}

Chemistry involving ro-vibrationally excited H$_2$ has long known to be significant for several key reactions in photon dominated regions (PDRs) of the interstellar medium  \citep[e.g.][]{Agundez10,Nagy13}. In these regions, the absorption of FUV photons in Lyman and Werner bands efficiently pumps H$_2$ in its excited electronic states. Approximately 10\% of the absorbed photons lead to H$_2$ dissociation, while 90\% radiatively de-excite to bound ro-vibrational levels of the ground electronic state. These excited ro-vibrational levels decay by means of quadrupole transitions which result in the emission of infrared photons \citep[e.g.][]{Black76}. The energy in these excited ro-vibrational levels is sufficient to overcome the reaction energy barrier of several reactions. Reactions of excited H$_2$ with C$^+$ and S$^+$ for instance are responsible for the formation of CH$^+$ and SH$^+$ ions in PDRs \citep[e.g.][]{Agundez10,Nagy13}. The efficiency of these reactions critically depends on the levels population of H$_2$. For C$^+$ for instance, the reaction with H$_2$ in ground vibrational state is endothermic by $\sim 4500$ K but becomes barrierless when reacting with H$_2$ in $v=1$ \citep{Zanchet13}. In the case of S$^+$, the reaction is  endothermic by $\sim 10000$ K and H$_2$ needs to be in  $v\gtrsim2$ to yield significant reactivity \citep{Zanchet19}.

Chemistry with ro-vibrationally excited H$_2$ in disks is often approximated by the so called pseudo-level approximation, in which all the excited vibrational levels of H$_2$ are grouped together into a single vibrational excited state H$_2^*$, corresponding to a pseudo-level $v^*=6$ and having a weighted averaged energy of $\sim 30000$ K \citep[see][and references therein]{Tielens85}. The use of this approximation has been shown to lead to satisfactory results when accounting for processes such as H$_2$ self-shielding and photodissociation as compared to detailed H$_2$ excitation calculations \citep[see Fig. 1 \& 2 of][for instance]{Draine96}. However, the use of this approximation to solve for chemistry with excited H$_2$ may lead to unrealistic results as the rates critically depends on the fractional populations in each ro-vibrational state. In particular, rate coefficients of species reacting with H$_2^*$ are often approximated using the rate for reaction with H$_2$ by reducing the exponential factor by the corresponding energy of the level \citep[i.e. $\sim 30000$K, see][]{Tielens85}. Recently, \citet{Cazzoletti18} used this approximation to explain the bright CN emission around the well studied disk TW Hya. They proposed that CN originates from the PDR surface by assuming a barrierless reaction of N with H$_2^*$, although there is a barrier of 12600K for it to react with ground state H$_2$. \citet{Visser18} used the same assumption to conclude that most of the observed HCN emission also comes from the PDR layer. \citet{Kamp17} follow a more conservative approach where the rate for reaction with H$_2^*$ is obtained by subtracting the energy corresponding to the first vibrational excited level $v=1$ for which $E_{vJ} \sim 6000$K. Full calculations accounting for the quantum ro-vibrational state of H$_2$ have been limited to few reactions such as H$_2$ with C$^+$, He$^+$, S$^+$, O, OH and CN, with important implications for the formation of molecules such as CH$^+$ and SH$^+$ \citep[][]{Agundez10,Agundez18}. The goal of this article is to explore the effects of the full vibrational level calculations of H$_2$ on the chemistry of protoplanetary disks. Modeling results are presented in Section \ref{sec:mod_res} and Section \ref{sec:conclusion} contains our conclusions.

\section{Modeling results}
\label{sec:mod_res}

In what follows, we consider the fiducial disk model of \citet{Ruaud19} to study the effects of H$_2$ excitation on the chemistry. The disk is assumed to have a mass $M_\text{disk} = 5 \times 10^{-3} M_\odot$  and a dust-to-gas mass ratio $\Sigma_d / \Sigma_g = 0.01$, and is constant with radius but it varies with height owing to dust settling \citep[see Appendix A of][for more details]{Ruaud19}.  Stellar parameters are assumed to be $M_* = 1M_\odot$, $R_* = 2.45 R_\odot$ and an X-ray luminosity of $10^{30}$ erg s$^{-1}$. The stellar spectrum is the TW Hya spectrum taken from the Leiden database \citep{Heays17}. 

\subsection{H$_2$ ro-vibrational excitation and level populations}

The model includes 100 ro-vibrational levels of H$_2$  in the ground electronic state. The level populations of H$_2$ are computed by including the effects of (1) radiative transitions, (2) collisional processes with He, H, H$^+$ and H$_2$, (3) FUV radiative pumping (4) X-ray excitation (5) excitation at H$_2$ formation and (6) the ortho/para conversion on grains. More details on the implementation of each of these process is given in the appendix. Even though X-ray pumping is taken into account, we find that it does not play a significant role in the non-thermal excitation of H$_2$ when compared to FUV pumping. This result is similar to earlier modeling results of \citet{Nomura07} where they study the relative roles of X-ray and UV excitation of H$_2$. Therefore, in what follows, we do not discuss the effects of X-ray pumping on chemistry any further.

\begin{figure*}
\centering
\includegraphics[width=17cm,trim=0 0.7cm 0 0, clip]{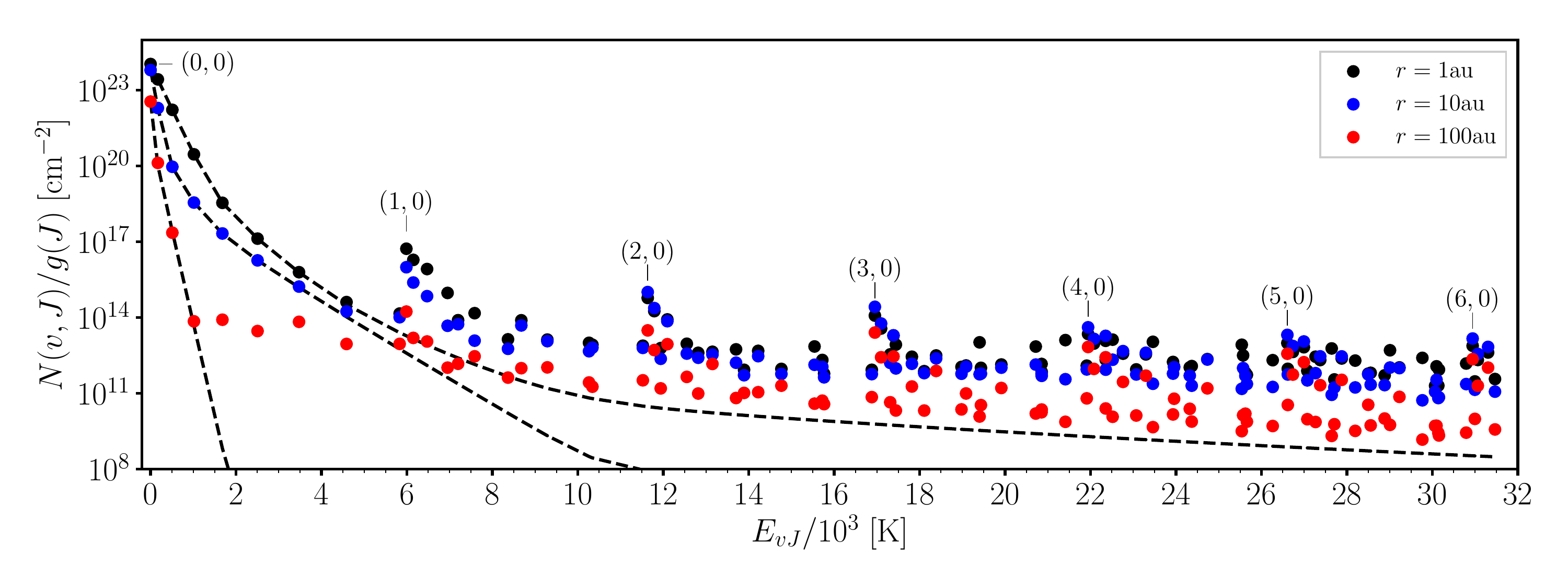}
\caption{\label{fig:exc_diag} Vertically integrated H$_2$ column density $N(v,J)$ divided by the level degeneracy $g(J)$ as a function of the energy of the level at $r=1$, 10 and 100 au from the star. Dashed lines are at LTE.}
\end{figure*}

Figure \ref{fig:exc_diag} shows the model calculated vertical column densities of all H$_2$ levels included in the model, divided by statistical weight, at $r=1$, 10 and 100 au from the star, and expected LTE values (dashed lines). For $r=1$ and 10 au, the lowest rotational levels of the ground vibrational level (i.e. $v=0$ and $J\leq8$) are at LTE and higher energy levels are mainly populated by FUV photon pumping. For $r=100$ au, only levels up to $J=3$ are thermally populated and departure from LTE induced by FUV photon pumping occurs at lower energy than for $r=1$ and 10 au. This is due to the the lower temperatures and densities of the outer disk as compared to the inner disk (i.e. radial temperature gradient). Rotational distributions in vibrationally excited states are characterized by rotational temperatures typical of fluorescent emission and ranging from $T_\text{rot} \sim$1000K and $\sim$2500K \citep[e.g.][]{Draine96}. Curvature observed in LTE values is explained by the vertical temperature gradient. Regions closer to the disk midplane, which dominate the contribution to the column of the lowest energy H$_2$ levels, are colder than regions located in the disk atmosphere. As a result, the LTE curve tends to flatten as the energy of the level increases.

\begin{figure*}
\centering
\includegraphics[width=17cm,trim=0 0.7cm 0 0.8cm, clip]{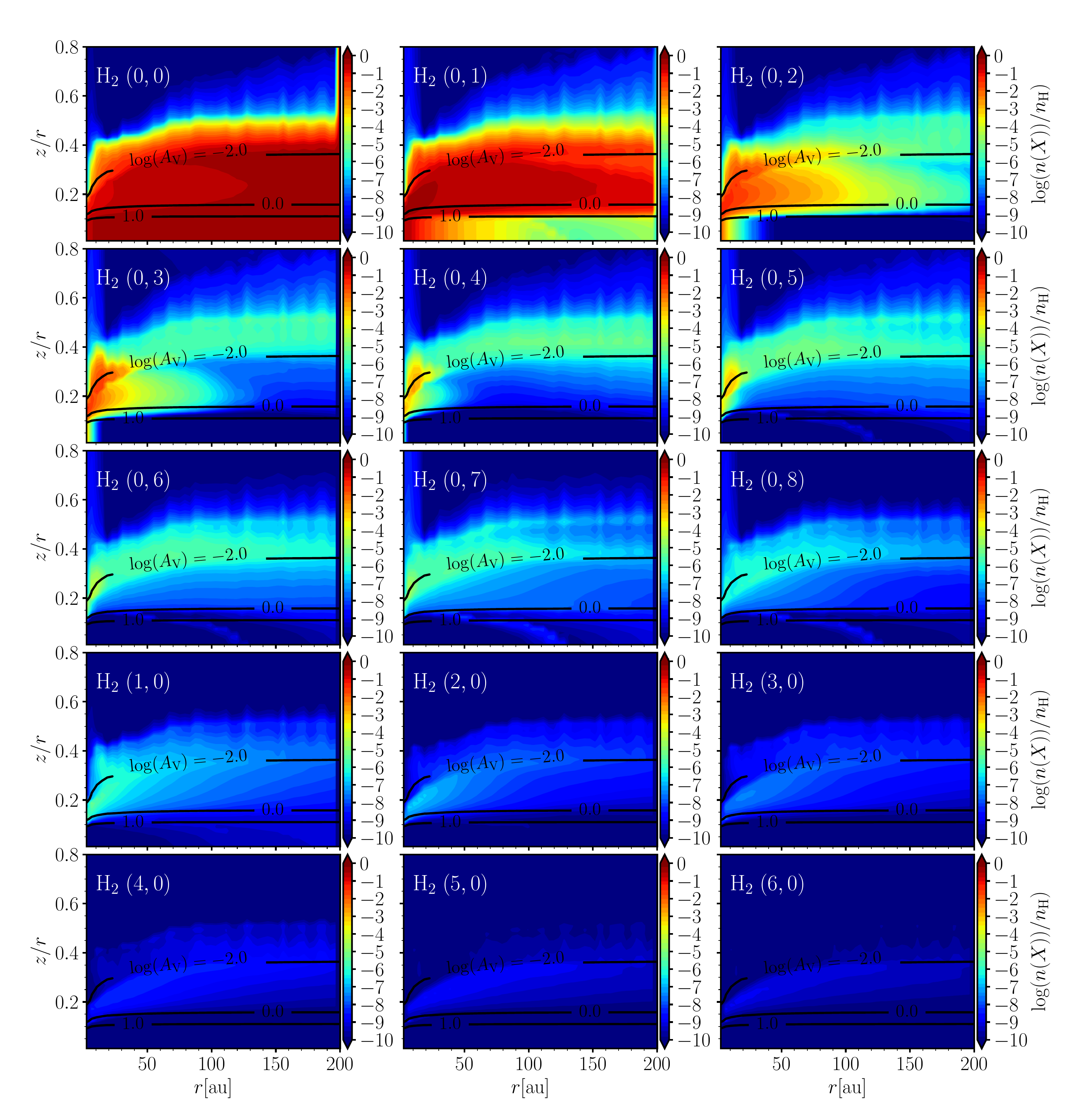}
\caption{\label{fig:h2_exc_map} Computed abundances of H$_2$ in different ro-vibrational levels as a function of the disk radius $r$ and normalized height $z/r$. Solid line contours show the $A_V$ to the star.}
\end{figure*}

Figure \ref{fig:h2_exc_map} shows the computed abundance maps of H$_2$ in different $(v,J)$ states. As seen in this figure (as well as Fig. \ref{fig:exc_diag}), most of the H$_2$ in the disk is in the ($v=0,J=0$) and ($v=0,J=1$) ground vibrational states and the abundance of H$_2$ in higher ro-vibrational states tends to decrease as the energy levels of the states increase. The abundance of the H$_2$ ($v=1,J=0$) state for instance reaches a peak abundance of few $10^{-6}$ with respect to the density n$_\text{H}$, while higher energy states have abundances $\lesssim 10^{-7}$. As will be discussed below, chemistry with excited H$_2$ critically depends on the fractional populations in each of these ro-vibrational states.

\subsection{Chemistry involving ro-vibrationally excitated H$_2$}

The effects of the full ro-vibrational level calculations of H$_2$ on the chemistry is explored for species having reaction barriers with ground state H$_2$. In the chemical network used here \citep[i.e. the KIDA network,][]{Wakelam15}, there are $\sim 80$ reactions, including important reactions with species such as C$^+$, S$^+$, N$^+$, C, O, N, and OH that mediate pathways to many other chemical species. Rate constants have been measured and/or calculated for some reactions with species such as C$^+$, S$^+$ and O. For $\text{C}^+(^2P) + \text{H}_2 \rightarrow \text{CH}^+ + \text{H}$, \citet{Gerlich87} experimentally derived rate constants for the reaction with H$_2(v=0,J=0-7)$ and for which we use analytical fits given in \citet{Agundez10}. 
For higher energy levels we use results from quantum calculations carried out by \citet{Zanchet13} and for which fits are available for H$_2(v=1,J=0)$ and H$_2(v=2,J=0)$ \citep[see Table 3 of][]{Zanchet13b}. For $\text{S}(^4S) + \text{H}_2 \rightarrow \text{SH}^+ + \text{H}$, we use rate constants given in \citet{Zanchet13b} for reaction with H$_2 (v=0,J=0)$ and $(v=1,J=0)$ states, and those given in \citet{Zanchet19} for reaction with H$_2 (v=2-5,J=0)$ states. In the case of $\text{O}(^3P) + \text{H}_2 \rightarrow \text{OH} + \text{H}$, we use fits to quantum calculations performed by \citep{Sultanov05} and for which expressions are given in \citet{Agundez10} for reaction with $\text{H}_2(v=0-3,J=0)$.
For these three reactions, since rate constants are available for $J=0$ only (i.e. except for $\text{C}^+(^2P) + \text{H}_2$ for which we have rates with H$_2(v=0,J=0-7)$), rate constants of higher $J$'s are assumed to be equal to the rate of the corresponding vibrational level in $J=0$. Moreover, for these reactions, rates for higher energy levels than the last available state are assumed to be equal to the rate in this state. For the remaining reactions having a barrier for reaction with H$_2$, reaction rates are approximated by

\begin{equation}
\label{eq:1}
k_{vJ}= \alpha (T/300)^\beta \exp (-\max[\gamma - E_{vJ},0]/T)~[\text{cm}^3 \text{s}^{-1}]
\end{equation}
where $\alpha$, $\beta$ and $\gamma$ are the usual rate constant parameters.
This expression assumes that the reaction proceeds without a barrier when the energy of the level overcomes the energy barrier of the reaction. Note that this assumption could be prone to error as rate constant enhancements are usually specific to each system and difficult to predict \citep[see also][]{Agundez10}. One such example is $\text{O}(^3P) + \text{H}_2$, which has an activation barrier of $\sim 4000$K, but for which an activation barrier persists even when H$_2$ is in the $v=3$ state, for which $E_{v} \sim 17000$K \citep{Sultanov05,Agundez10}. 

\begin{figure}
\centering
\includegraphics[width=8cm,trim=0 0.7cm 0 0.8cm, clip]{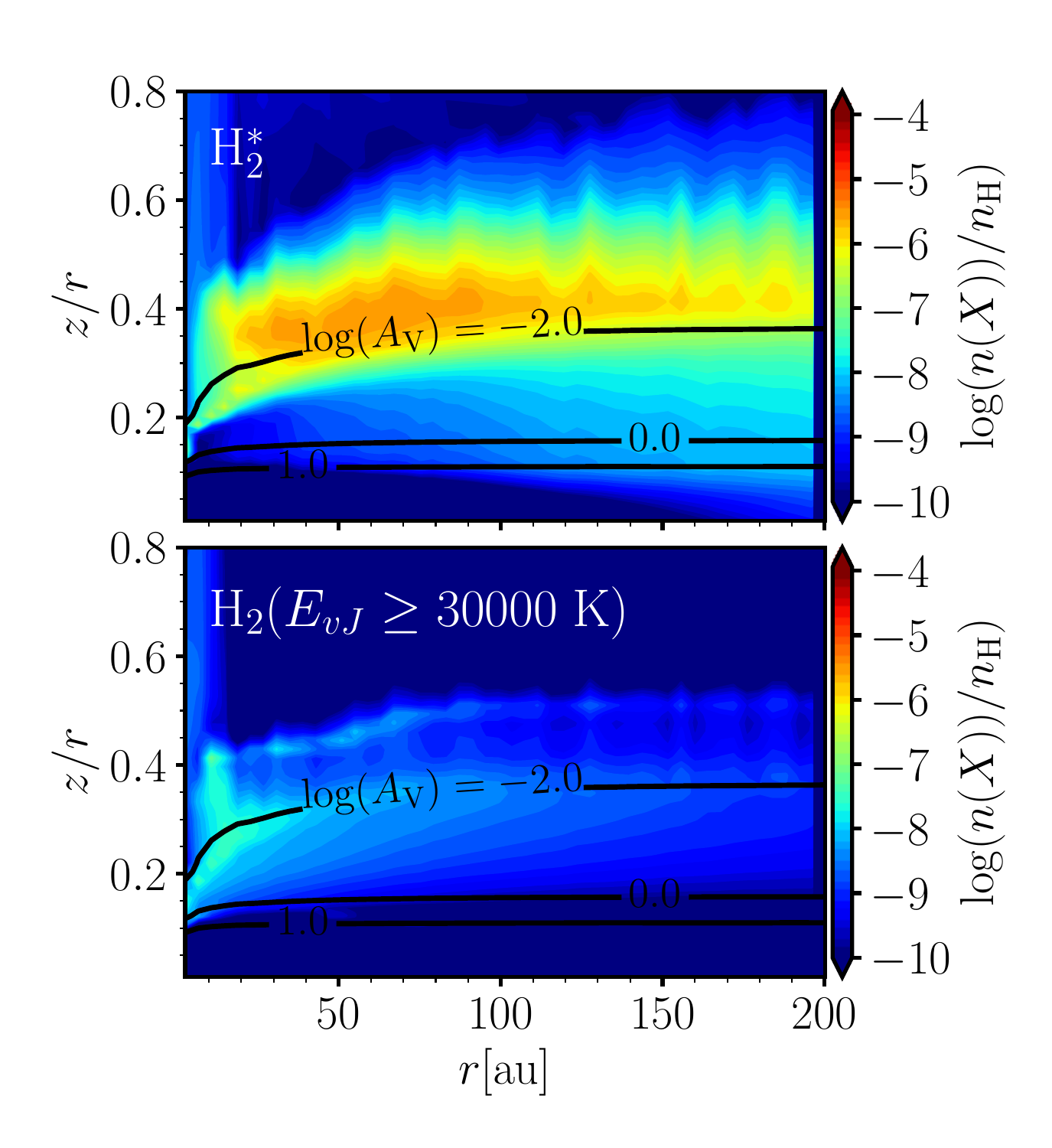}
\caption{\label{fig:h2v}  Computed abundance of excited H$_2$ as a function of the disk radius $r$ and normalized height $z/r$. The top figure shows result obtained in the case of the pseudo-level approximation and the bottom figure shows results obtained with the full model and for levels with $E_{vJ} \geq 30000$K.}
\end{figure}

Three types of models are considered in the following: (1) a reference model in which the effect of the chemistry with ro-vibrationally excited H$_2$ is neglected\footnote{It should however be noted that reaction parameters that are determined experimentally have a temperature dependence that implicitly consider the excitation of H$_2$.}, (2) a model in which all the excited vibrational levels of H$_2$ are grouped together into a single vibrational excited state having a weighted averaged energy of 30160 K corresponding to H$_2(v=6)$ (the so-called pseudo-level approximation), and (3) a model where the full H$_2(v,J)$ population calculation is taken into account (see Table \ref{tab:mod}). 
For model M2, we follow the treatment described in \citet{Tielens85} in which H$_2^*$ is formed by FUV pumping and destroyed by radiative decay and collisional de-excitation by H and H$_2$. The FUV pumping rate is assumed to be 9 times the H$_2$ photodissociation rate and the radiative decay rate is $2\times 10^{-7}$ s$^{-1}$. For collisional de-excitation by H and H$_2$ we use rates given in Equation (A14) of \citet{Tielens85}. These rates have recently been updated by \citet{Visser18} based on more recent calculations. A comparison with the rates used in \citet{Visser18} however shows that the impact of these updated rates is relatively small. This is due to the fact that radiative decay dominates H$_2^*$ destruction in most of the disk. The only location where the use of these updated rates leads to differences is the inner disk  (i.e. $r\lesssim20$ au) where higher densities result in collisional de-excitation dominating radiative decay. Finally, reaction rates with H$_2^*$ in this model are computed using Eq. \ref{eq:1} and  assuming $E_{v^*} = 30160$K.  A comparison between the computed abundance map of H$_2^*$ calculated with the pseudo-level approximation and for all levels with $E_{vJ} \geq 30000$K of the full model is shown in Fig. \ref{fig:h2v}. 

\begin{figure*}
\centering
\includegraphics[width=17cm,trim=0 2.2cm 0 0.8cm, clip]{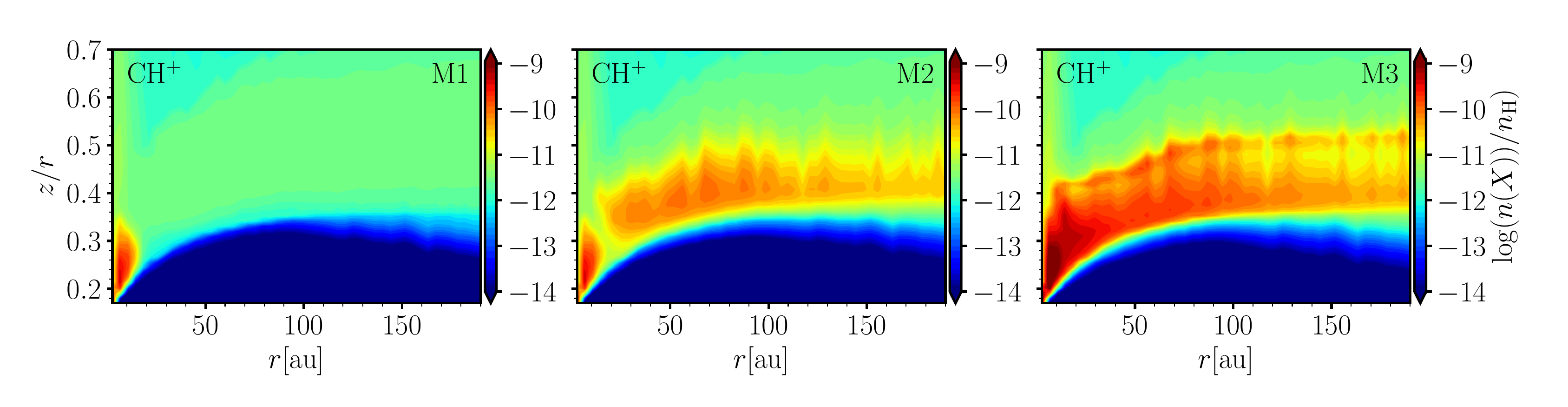}\\
\includegraphics[width=17cm,trim=0 2.2cm 0 0.8cm, clip]{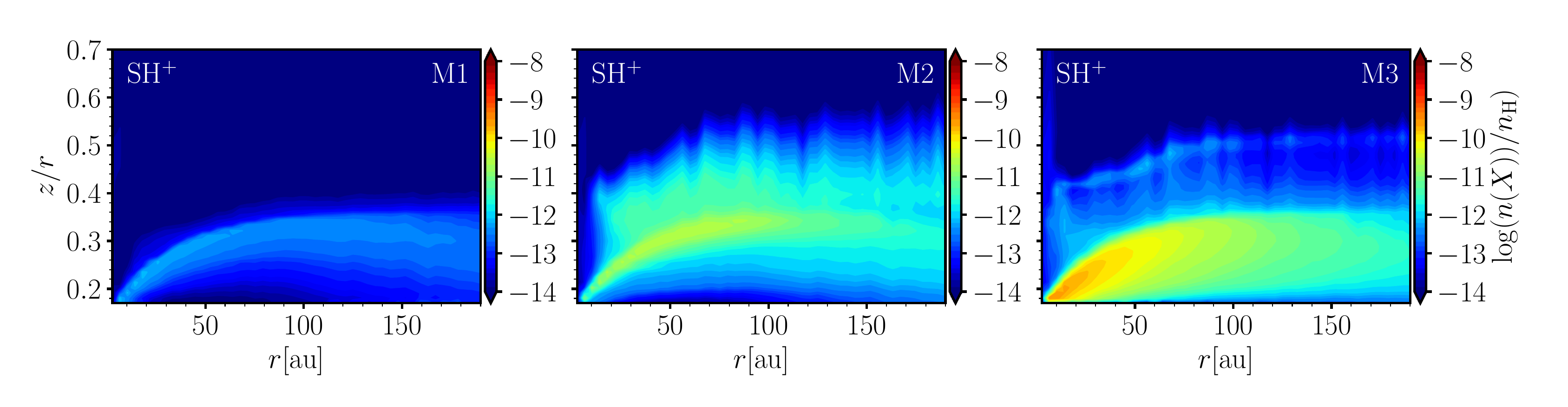}\\
\includegraphics[width=17cm,trim=0 0.7cm 0 0.8cm, clip]{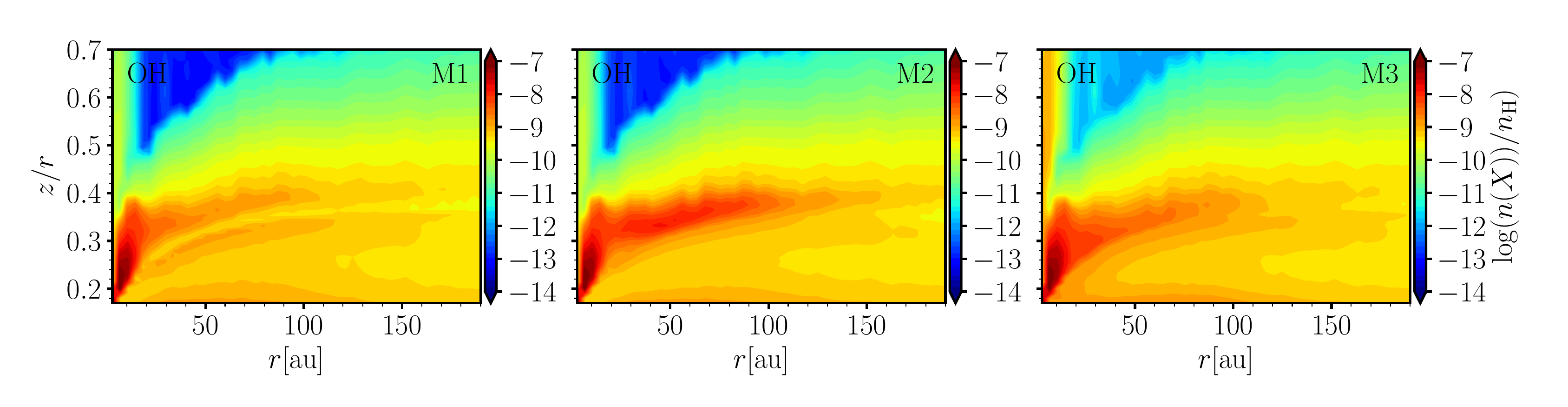}
\caption{\label{fig:res1} Computed abundance map of CH$^+$, SH$^+$ and OH as a function of the disk radius $r$ and normalized height $z/r$ for model M1, M2 and M3 (see Table \ref{tab:mod}).}
\end{figure*}

\begin{figure*}
\centering
\includegraphics[width=17cm,trim=0 2.2cm 0 0.8cm, clip]{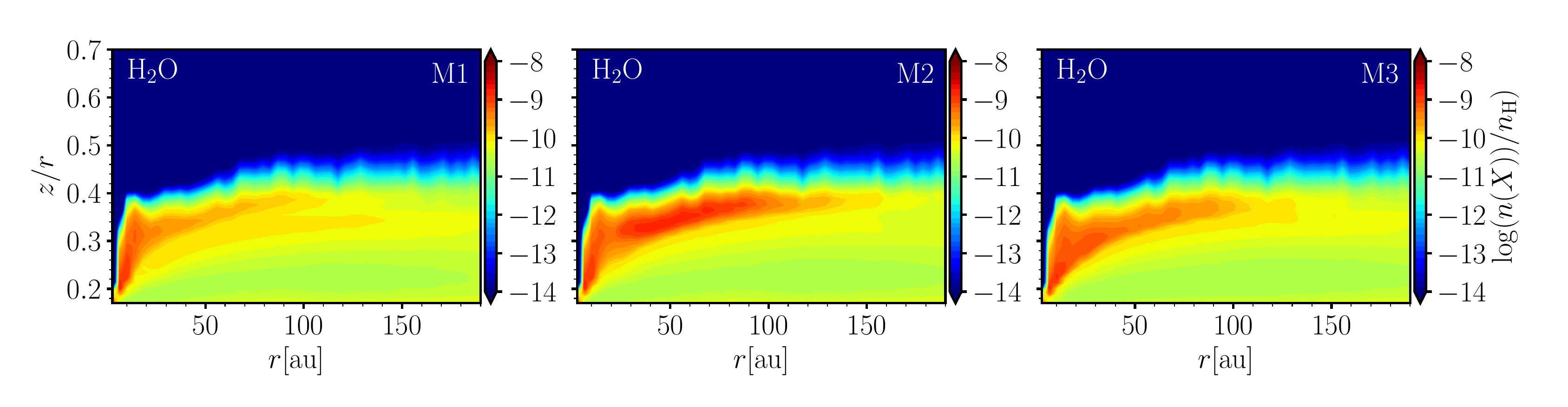}\\
\includegraphics[width=17cm,trim=0 2.2cm 0 0.8cm, clip]{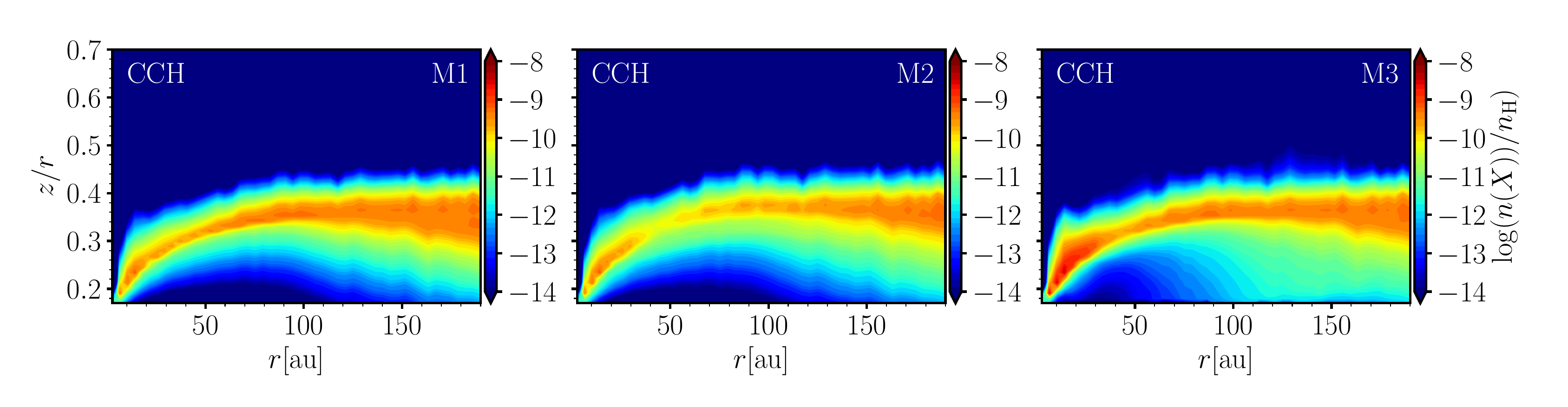}\\
\includegraphics[width=17cm,trim=0 2.2cm 0 0.8cm, clip]{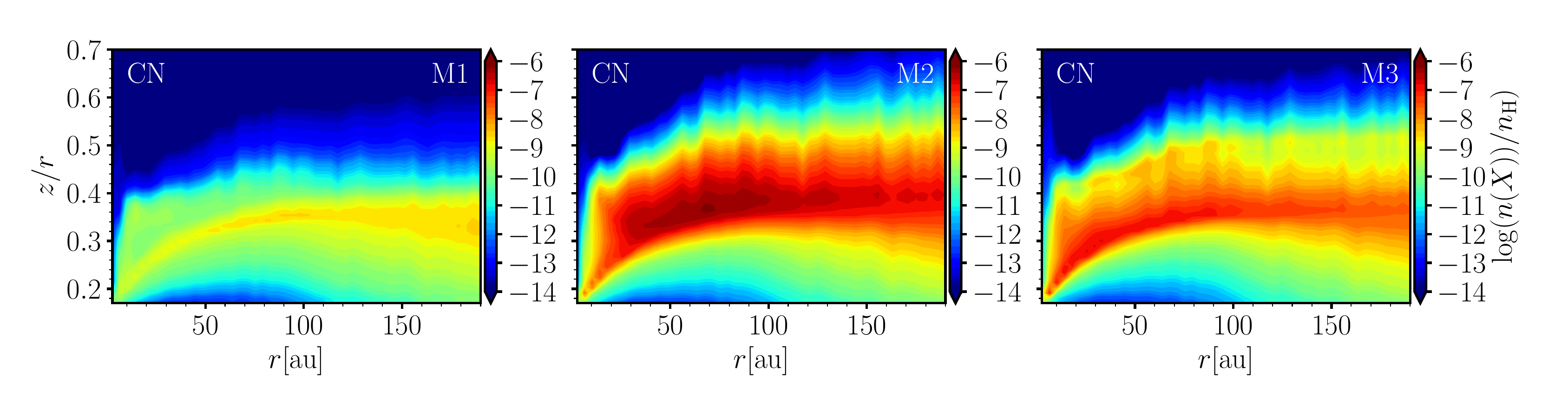}\\
\includegraphics[width=17cm,trim=0 2.2cm 0 0.8cm, clip]{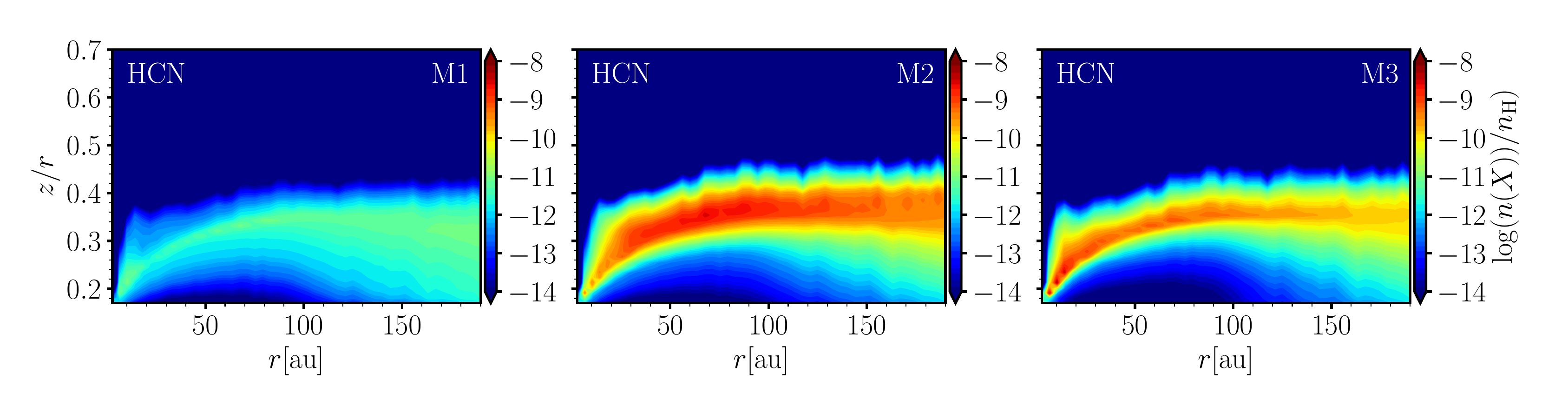}\\
\includegraphics[width=17cm,trim=0 2.2cm 0 0.8cm, clip]{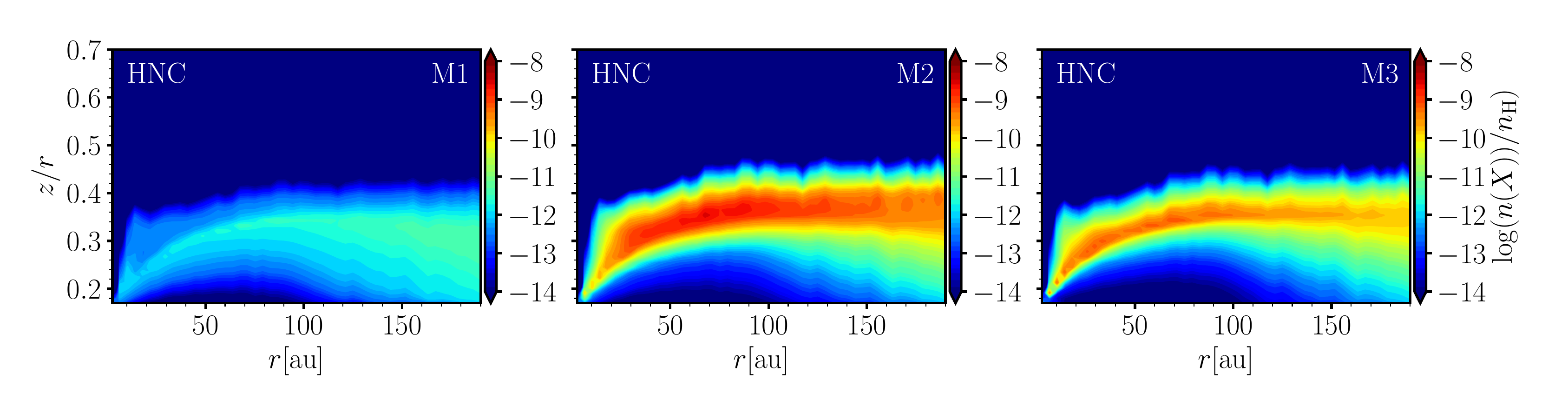}\\
\includegraphics[width=17cm,trim=0 0.7cm 0 0.8cm, clip]{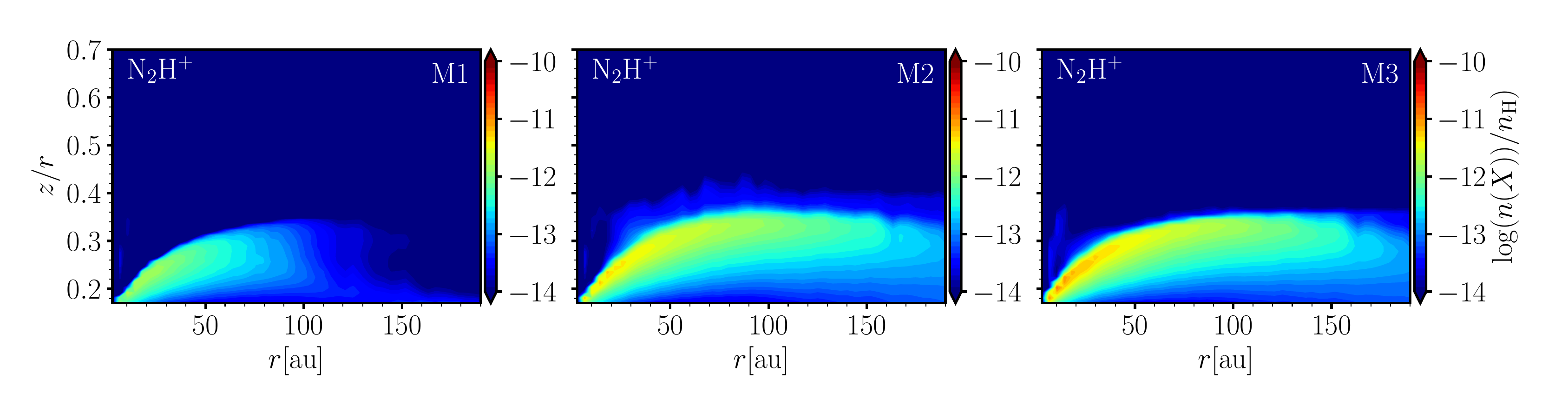}
\caption{\label{fig:res2} Same as Fig. \ref{fig:res1} but for H$_2$O, C$_2$H, CN, HCN, HNC and N$_2$H$^+$.}
\end{figure*}

\begin{figure*}
\centering
\includegraphics[width=15cm,trim=0 0.7cm 0 0.5cm,clip]{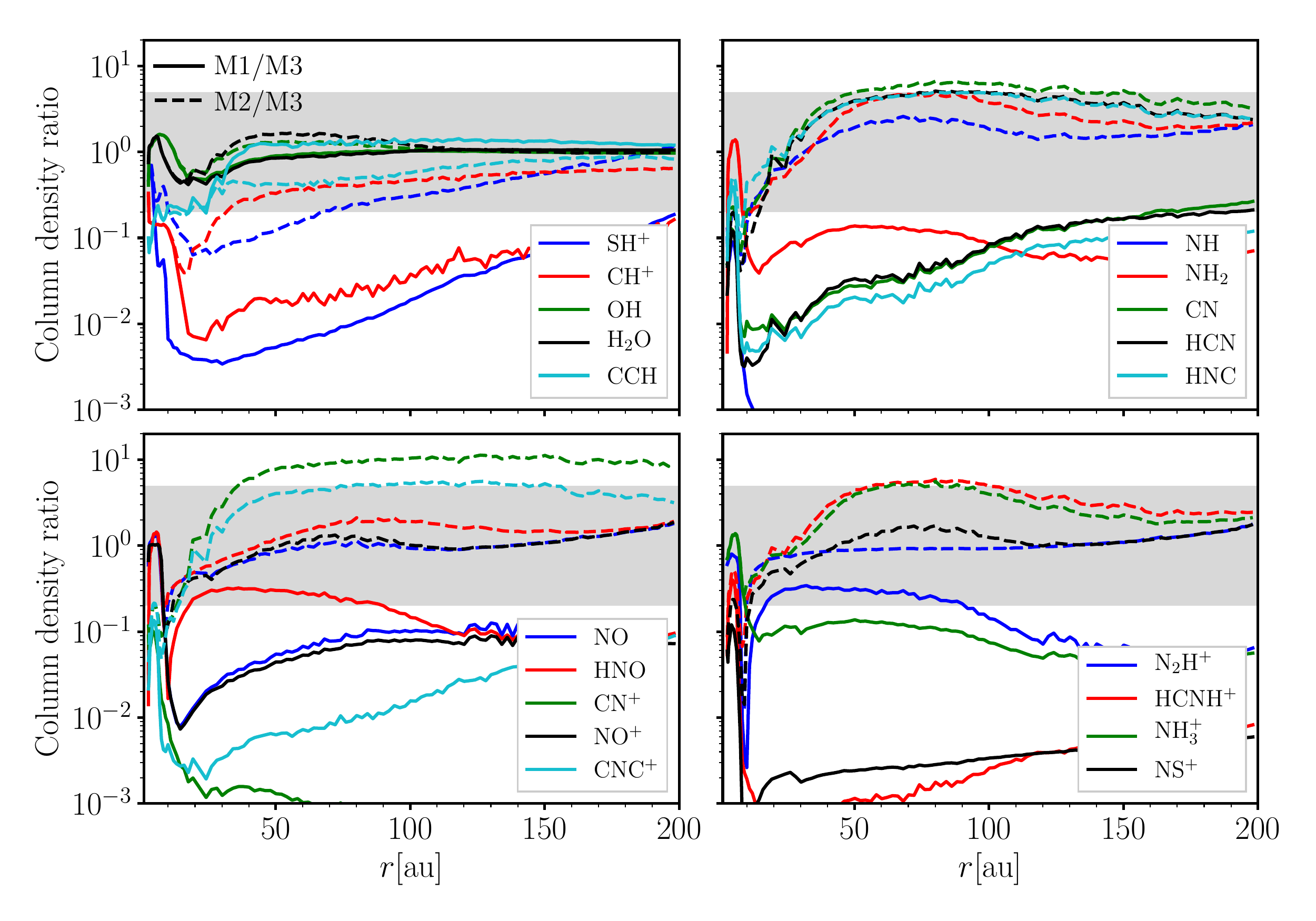}
\caption{\label{fig:colum_densities_ratios} Vertical column density ratios of a selection of gas-phase molecules as a function of the distance from the star. Solid lines correspond to ratios between column densities computed in model M1 (no chemistry with excited H$_2$) and M3 (full model) while dashed lines show ratios between results obtained from model M2 (pseudo-level approximation) and M3. The gray shaded area indicate a factor of 5 deviation compared to results obtained from the full model.}
\end{figure*}

\begin{table}
    \centering
    \caption{\label{tab:mod} Model description.}
    \begin{tabular}{l l}
    \hline
    \hline
   	 	& Model description\\
    \hline
	M1	& Reference model in which the chemistry\\
		& of excited H$_2$ is neglected \\
	M2 	& Chemistry of excited H$_2$ assuming \\
		& the pseudo-level approximation \\
	M3	& Chemistry of excited H$_2$ with \\
		& full H$_2$ population calculation \\
    \hline
    \end{tabular}
\end{table}
In the following, the chemistry in regions with $z/r<0.2$, which trace regions where grain surface chemistry is important, is not presented. Readers interested in the chemistry of these regions can refer to papers such as \citet{Semenov11}, \citet{Walsh14} and \citet{Ruaud19}, for detailed models of gas-grain chemistry using large surface reaction networks. Figs. \ref{fig:res1} and \ref{fig:res2} show the computed abundance maps of CH$^+$, SH$^+$, OH, H$_2$O, C$_2$H, CN, HCN, HNC and N$_2$H$^+$ computed in each of these models  and show that the chemistry with ro-vibrationally excited H$_2$ is important in the surface layers of the disk (i.e. in regions with $z/r>0.2$, where H$_2$ ro-vibrational levels are efficiently populated by the absorption of FUV photons, see Fig. \ref{fig:h2_exc_map}). The use of the pseudo-level approximation as compared to the full H$_2$ population shows important differences for species such as CN, HCN  and HNC for which the computed abundances in the PDR layer are overestimated by approximately one order of magnitude in the pseudo-level approximation model (M2) as compared to the full H$_2$ population model (M3). These differences are also seen in Fig.  \ref{fig:colum_densities_ratios} which shows vertical column density ratios for models M1 and M2 with respect to the full calculation M3 of some abundant gas-phase species. Details on some important reaction pathways are discussed below.

\subsubsection{Carbon chemistry}

The inclusion of C$^+$ + H$_2(v,J)$ has an important impact on the formation of CH$^+$ as compared to the reference model M1. At $r>20$ au, CH$^+$ abundance is increased by almost two orders of magnitude in the M2 and M3 models as compared to M1, while the M3 model has $\sim3$ times higher CH$^+$ than the M2 model. In the full calculation model, CH$^+$ is mainly formed by reaction with H$_2$ in the $(v=0,J=8)$, $(v=1,J=0)$ and $(v=1,J=1)$ state, for which $E_{vJ}\gtrsim 4640$ K. Enhanced formation of CH$^+$ in the irradiated layer is in agreement with \citet{Agundez18}, even though our computed abundances are lower by a factor of $\sim 10$. This could be the result of different disk physical parameters and/or the use of a different stellar spectrum. As seen in Figs. \ref{fig:res2} and \ref{fig:colum_densities_ratios}, the enhanced formation of CH$^+$ only has a moderate impact on the formation of C$_2$H, a bright molecule at millimeter wavelengths in disks for which most models fail to reproduce observed emission without invoking enhanced C/O ratios \citep[e.g.][]{Bergin16,Cleeves18,Miotello19}. In all three models, C$_2$H formation starts with CH + C$^+$ which forms C$_2^+$. C$_2^+$ then reacts with H$_2$ to form successively C$_2$H$^+$ and C$_2$H$_2^+$, which then recombine to form C$_2$H. The key molecule in this sequence is CH whose formation is primarily initiated by the radiative association of C$^+$ and H$_2$ to form CH$_2^+$.  CH$_2^+$ then reacts with H$_2$ to form CH$_3^+$, which then dissociatively recombine to form CH. In model M2 and M3, the reaction of C with excited H$_2$ also contributes to the formation of CH. In model M3, enhanced formation of CH$^+$ plays a role on the formation of C$_2$H at r$\lesssim 40$ au, where C$_2$H abundance is enhanced by a factor of $\sim 5$ when compared to model M1. In this region, the enhanced formation of CH$^+$ contributes to the formation of CH through CH$^+$ + H$_2$ which forms CH$_2^+$ and then CH through the sequence described above. In model M2, the decreased abundance of C$_2$H at $30\lesssim r \lesssim 100$ au is found to be linked to the enhanced production of CN in this region which efficiently competes with C$_2$H formation. It should however be noted that the enhanced formation of C$_2$H due to chemistry with excited H$_2$ may not be enough to explain observed C$_2$H emission using ISM-like C/O ratios and C/O$\gtrsim 1$ may still be required. Increased vertical column densities of at least an order of magnitude seems to be needed in order to reproduce C$_2$H observations as compared to what is predicted with standards C/O ratios \citep[e.g.][]{Bergin16,Cleeves18,Miotello19}. It is however still unclear whether this change in disks C/O ratios is induced by the differential sequestration of oxygen and carbon at the surface of large grains in the disk miplane \citep[e.g.][]{Krijt18,Krijt20} or by photodestruction of carbon grains in disks atmosphere \citep[e.g.][]{Anderson17,Bosman21}.

\subsubsection{Oxygen chemistry}
The inclusion of O + H$_2(v,J)$ has only a moderate impact on the computed abundances of OH and water in the M3 model as compared to M1 (see Fig. \ref{fig:res1} and \ref{fig:res2}). In both models, OH and water formation are mainly driven by X-rays through the production of H$^+$ and H$_3^+$, and their successive reaction with atomic oxygen. As seen in these figures, M2 overestimates slightly the abundance of OH and water (i.e. by a factor of $\sim 5$) in a thin layer located at $z/r \sim 0.3 - 0.4$ and extending from $r=20$ and 100 au as compared to M3. The higher abundance of OH and H$_2$O in this layer are found to be linked to the enhanced production of CN in this model and in particular a slightly lower abundance of C$^+$ which is one of the main destruction channels of OH and water in this layer. The chemistry of other molecules such as NO, HNO and NO$^+$ are also found to be affected by chemistry with ro-vibrationally excited H$_2$ (see Fig. \ref{fig:colum_densities_ratios}). For these molecules, this is due to the enhanced formation of NH and NH$_2$ (see Section \ref{sec:n_chem}) and their subsequent reaction with O.

\subsubsection{Sulfur chemistry}
Even though the abundance of SH$^+$ is low throughout the disk, the inclusion of S$^+$ + H$_2(v,J)$ has an important impact on its formation (see Fig. \ref{fig:res1}). As seen in this figure, models M2 and M3 give relatively different results with SH$^+$ being overestimated in the vertical directions of model M2 and slightly underestimated in the inner part of the disk (i.e. at $r\lesssim 100$ au). Quantum calculations show that H$_2$ needs to be in $v\geq2$ states for the reaction to proceed without a barrier  \citep{Zanchet19} and its efficiency thus depends on the populations in the H$_2(v\geq 2)$ states. In model M3, H$_2(v\geq 2)$ states are more efficiently populated in the inner disk (the abundance of H$_2(v\geq 2)$ is on the order of few $10^{-8}-10^{-7}$ up to $r\sim 100$ au, see Fig. \ref{fig:h2_exc_map}). At  greater distance from the star, the populations in these states decrease significantly hence lowering the formation efficiency of SH$^+$. Except for SH$^+$, sulfur chemistry is not particularly affected by chemistry with excited H$_2$.

\subsubsection{Nitrogen chemistry}
\label{sec:n_chem}
Nitrogen chemistry is strongly affected by the chemistry with ro-vibrationally excited H$_2$. In particular, the abundance of species such as CN, HCN and HNC is increased by a factor of $\sim 100$ in model M3 as compared to the reference model M1 (see Fig. \ref{fig:res2}). For model M2, the abundance of these species is increased by almost a factor of $\sim 10$ as compared to model M3 at the peak abundance (and a factor of 3 to 5 in terms of the computed vertical column densities, see Fig. \ref{fig:colum_densities_ratios}). In both model M2 and M3, the formation of CN and HCN/HNC starts from N + H$_2(v,J)$ to form NH for which the reaction barrier is 12650 K. NH then reacts with C$^+$ to form CN$^+$. CN is then formed by charge exchange with H while HCN/HNC are formed through reaction of CN$^+$ with H$_2$ to form successively HCN$^+$ and HCNH$^+$, which then recombines with electrons to form HCN and HNC. The overestimation of CN and HCN/HNC in model M2 as compared to model M3 comes from the fact that the fractional population of H$_2$ in $(v>2)$ states, for which $E(v,J) > 11600$ K, are relatively low as compared to what is predicted for H$_2^*$ (see Fig. \ref{fig:h2_exc_map} and \ref{fig:h2v}). As seen in Fig. \ref{fig:res2} and \ref{fig:colum_densities_ratios}, the increased formation of CN and HCN also impacts N$_2$H$^+$ for which the computed vertical column densities are increased by a factor of at least 3 in models M2 and M3 as compared to model M1. The increased production of NH is also found to impact the formation of molecules such as NH$_2$, NO, HNO, NO$^+$, CN$^+$, HCNH$^+$, CNC$^+$, NS$^+$, SiN and SiN$^+$ (see Fig. \ref{fig:colum_densities_ratios}). Recently, \citet{Cazzoletti18} used the pseudo-level approximation to explain CN emission around the well studied disk TW Hya as originating from the PDR surface and being formed by the same reaction of N with excited H$_2$, where they assumed no barrier (i.e. similar to our model M2). \citet{Visser18} used the same assumption to conclude that most of the observed HCN emission also comes from the PDR layer. While the impact of this reaction on the formation of CN and HCN is confirmed by models shown here, it is however found that the use of the pseudo-level approximation lead to a factor of 3 to 5 overestimation in their computed vertical column densities (see Fig. \ref{fig:colum_densities_ratios}).

\subsection{Sensitivity to the stellar FUV photons flux}
\begin{figure*}
\centering
\includegraphics[width=17cm,trim=0 0.7cm 0 0.5cm,clip]{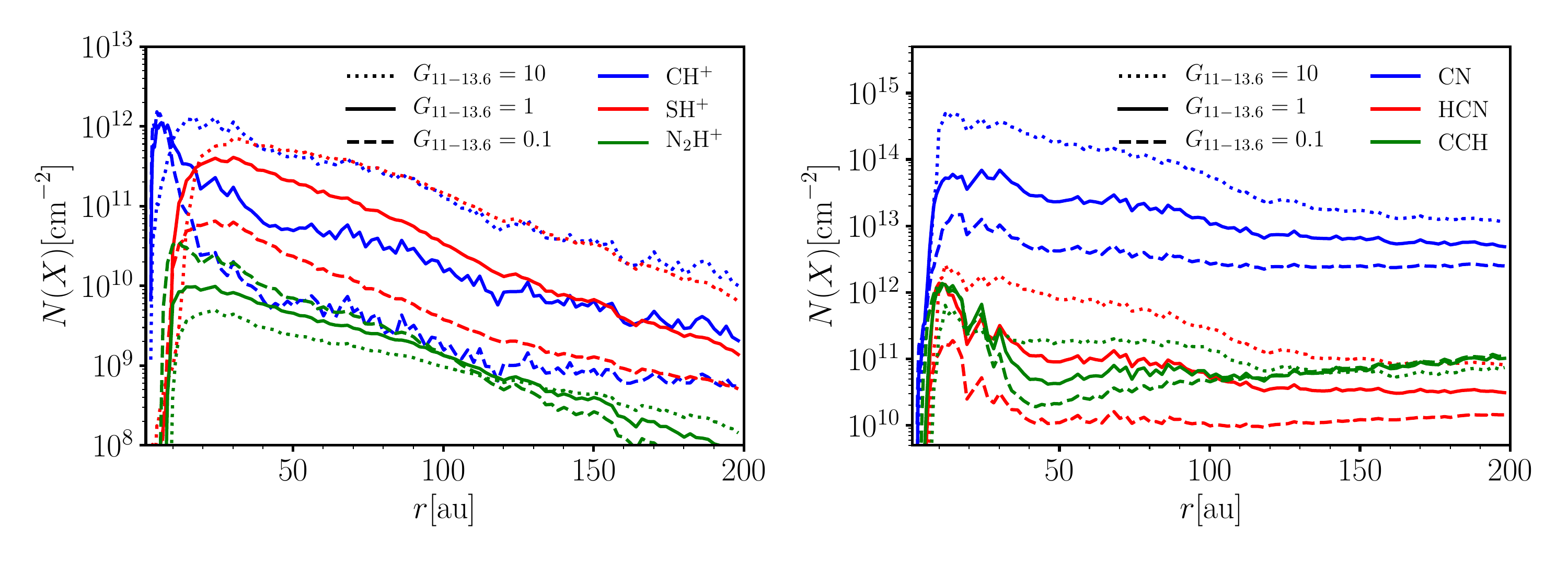}
\caption{\label{fig:col_dens}
Computed vertical column densities of CH$^+$, SH$^+$, N$_2$H$^+$, C$_2$H, CN and HCN as a function of the disk radius $r$ for model M3 and where the FUV flux in the 11 to 13.6eV photons range was varied by a factor of 10. $G_{11-13.6}=1$ used for the reference model corresponds to a photon flux of $4\times 10^{13}$ photons cm$^{-2}$ s$^{-1}$ in the 11-13.6 eV range and at 1 au from the star.}
\end{figure*}

In this section we study the effect of the FUV flux on the modeling results. As seen previously, the chemistry of excited H$_2$ sensitively depends on the fractional populations in each ro-vibrational state of H$_2$ which in turn depend on FUV pumping. We focus on the 11 to 13.6eV photons range which is poorly constrained in disks \citep[e.g.][]{France14} but yet fundamental for H$_2$ excitation and disk photochemistry. In particular, it is important to note that most H$_2$ fluorescent lines only depend on the photon flux in this energy range. This is because FUV pumping tends to be dominated by electronic transitions from the lower vibrational levels of the ground state $X ^1\Sigma_g^+$, (e.g. electronic Lyman and Werner transitions from H$_2(v=0,J<8)$ all lie at $\lambda <1150$\AA). In the following, we consider two variations of model M3 in which we vary the FUV flux in the 11 to 13.6 eV range by a factor of 10. It should however be noted that the disk physical structure is kept the same and that only the chemistry is recomputed in these models. Fig. \ref{fig:col_dens} compares the effect of varying the FUV flux on the computed vertical column density of a selection of gas-phase molecules which are strongly affected by a change in the FUV flux. In particular, the computed vertical column densities of CH$^+$, SH$^+$, CN and HCN vary by a factor of $\sim 10$ when the FUV flux in the 11-13.6 eV range varies by a factor of 10. C$_2$H is found to be relatively insensitive to the FUV flux, except in a small region located at $30\lesssim r \lesssim 100$ au where its computed vertical column density varies by at most a factor of $\sim 5$ when compared to the reference model (i.e.  $G_{11-13.6}=1$). Increased CN, HCN and C$_2$H column densities with FUV flux could explain some of the positive correlations found in the emission of these molecules \citep[e.g.][]{Bergner19}. 

\section{Conclusion}
\label{sec:conclusion}

The effect of H$_2$ ro-vibrational excitation on the chemistry of protoplanetary disks is studied using a framework that solves for the disk physical and chemical structure and includes a detailed calculation of H$_2$ level populations.
Chemistry with ro-vibrationally excited H$_2$ is found to be efficient in the irradiated surface layer of disks where H$_2$ is efficiently pumped by FUV photons and its efficiency sensitively depends on the fractional populations in each ro-vibrational state of H$_2$. Chemistry with excited H$_2$ is found to play an important role on the formation of several commonly observed species in disks and in particular CN, HCN and HNC, for which the computed vertical column densities are increased by 1 to 2 orders of magnitude for the full H$_2$ population calculation compared to calculations with no treatment of ro-vibrationally excited H$_2$, and by factors of $\sim3-5$ when compared to the commonly used pseudo-level approximation where all excited H$_2$ is considered to be at a single energy.  Chemistry with ro-vibrationally excited H$_2$ also enhances the formation of molecules such as SH$^+$, CH$^+$, N$_2$H$^+$, NH, NH$_2$, NO, HNO, NO$^+$, CN$^+$, HCNH$^+$, CNC$^+$, NS$^+$, SiN and SiN$^+$. For most of these molecules, except CH$^+$ and SH$^+$, the key reaction is N + H$_2\rightarrow $ NH which has a barrier of 12650 K with ground state H$_2$. Even though few studies exist on this particular system \citep[e.g][]{Pascual02}, rate constants as a function of H$_2$ ro-vibrational levels are missing. Here it was assumed that the reaction proceeds without a barrier when the energy of the level overcomes the energy barrier of the reaction (see Eq. \ref{eq:1}). However this could be a strong over simplification of the problem and rate constants should therefore be obtained for a more accurate assessment of CN/HCN chemistry. This is even more critical that molecules such as C$_2$H, CN or HCN are now commonly used to infer changes in the elemental disk C/O and N/O ratios \citep[e.g.][]{Bergin16,Cleeves18,Miotello19}, with implications for planetesimal formation and the composition of exoplanet atmospheres.
It is also shown that the formation of these molecules is greatly affected by a change in the FUV flux in the 11 to 13.6eV energy range. In particular we find a positive correlation between the increase in the FUV flux and the computed vertical column densities of molecules such as CN, HCN and to a lesser extent C$_2$H. However, the strength of the FUV photon flux at energies that pump H$_2$ in protoplanetary disks (i.e. mostly 11-13.6 eV) is not well constrained and this further impacts our ability to make inferences on the chemical state of the disk during planet formation epochs. With the upcoming launch of JWST, it will soon be possible to probe excited H$_2$ which will help to further constrain disks chemistry and potential molecular tracers of planet formation processes. Finally, it is found that computed vertical column densities from the common pseudo-level approximation for H$_2$ chemistry for most species agree within a factor $\lesssim 5$  with the more detailed calculations presented here.

\acknowledgments
I thank O. Roncero and A. Zanchet for sharing their computational results on state-to-state chemistry and for fruitful discussions around these data. I acknowledge support by the National Aeronautics and Space Administration through the NASA Astrobiology Institute under Cooperative Agreement Notice NNH13ZDA017C issued through the Science Mission Directorate.

\appendix

\section{Radiative transitions}
\label{sec:rad_trans}

Due to its homonuclearity and symmetry, H$_2$ has no permanent dipole moment. As a result, electric dipole ro-vibrational transitions of H$_2$ are forbidden and only weak quadrupole transitions are allowed. For ro-vibrational levels within the ground electronic state we use energy levels compiled in \citet{Dabrowski84} and quadrupole radiative transition probabilities from \citet{Wolniewicz98}. For radiative transitions of the Lyman and Werner band systems (i.e. $B^1\Sigma_u \rightarrow X^1\Sigma^+_g$ and $C^1\Pi_u \rightarrow X^1\Sigma^+_g$), we use energy levels and transition probabilities calculated by \citet{Abgrall93a,Abgrall93b,Abgrall00}.

\section{Collisional processes}
\label{sec:coll_trans}

Collision excitation are included for collisions with H \citep{Wrathmall07}, He \citep{Flower98}, H$_2$ \citep{Flower98a} and H$^+$  \citep{Gerlich90}. For collisions with He and H$_2$ we use fit to collisional de-excitation rates provided by \citet{LeBourlot99} in the range $100\lesssim T_\mathrm{gas} \lesssim 6000$ K. Collisional excitation rates are then computed using the detail balance equation
\begin{equation}
q(v'J'\rightarrow vJ) = \frac{g_J}{g_J'}  q(vJ\rightarrow v'J') \exp\Bigg(-\frac{E_{vJ} - E_{v'J'}}{T}\Bigg)
\end{equation}
where $g_J$ is the statistical weight of the level $J$ and $E_{vJ}$ its energy. Statistical weights are given by $g_J = (2J+1)$ for para-H$_2$ and $g_J = 3(2J+1)$ for ortho-H$_2$.

\section{FUV radiative pumping}
\label{sec:fuv_pump}

The FUV pumping rate from a ro-vibrational level $(v,J)$ of the ground electronic sate to level $(v^*,J^*)$ of an upper electronic state is given by \citep[e.g][]{vanDishoeck86,Sternberg89}

\begin{equation} 
\label{eq:B1}
P(vJ\rightarrow v^*J^*) = \int I_\nu\sigma_\nu(vJ\rightarrow v^*J^*)\exp\big( - N_{vJ} \sigma_\nu(vJ\rightarrow v^*J^*) \big) d\nu
\end{equation}

where $\nu$ is the transition frequency, $I_\nu$ is the FUV photon intensity (in photons.cm$^{-2}$.s$^{-1}$.Hz$^{-1}$), $\sigma_\nu$ is the photon absorption cross section of the transition, and $N_{vJ}$ is the column density of H$_2$ molecule in level $(v,J)$. Assuming that the FUV intensity is constant over the line profile and defining the equivalent width of the absorption line

\begin{equation}
W(vJ\rightarrow v^*J^*) = \int \Big[1 - \exp\big( - N_{vJ} \sigma_\nu(vJ\rightarrow v^*J^*) \big) \Big] d\nu
\end{equation}

as well as the line self-shielding function

\begin{equation}
\theta(vJ\rightarrow v^*J^*) = \Bigg(\frac{\pi e^2}{m_e c}f_\text{osc}(vJ\rightarrow v^*J^*)\Bigg)^{-1} \frac{dW(vJ\rightarrow v^*J^*)}{dN_{vJ}}
\end{equation}

equation \ref{eq:B1} may be rewritten

\begin{equation}
P(vJ\rightarrow v^*J^*) = \frac{\pi e^2}{m_e c} I_\nu f_\text{osc}(vJ\rightarrow v^*J^*)\theta(vJ\rightarrow v^*J^*)
\end{equation}

where $f_\text{osc}(vJ\rightarrow v^*J^*)$ is the oscillator strength of the transition. The line self-shielding function is calculated following the analytical treatment of \citet{Federman79}.

\section{Excitation by X-rays}
\label{sec:xray_pump}
Fast electrons created by X-rays ionization can excite H$_2$ ro-vibrational levels via collisions. This can occur either by direct collisional excitation within the ground vibrational level or by cascade following excitation of the electronic states \citep[e.g.][]{Gredel95,Tine97}. \citet{Tine97} provide entry probabilities for excitation rates by X-rays for H$_2$ in the ground vibrational state to H$_2(v=0-14,J=0-11)$ and use tabulated entry probabilities obtained for a fractional ionization of $10^{-4}$. 

\section{Excitation at formation and o/p conversion on grains}
\label{sec:grain_exc}
For the excitation at formation, we follow an approach similar to \citet{LePetit06} and consider an equipartition of the energy released at formation (i.e. 4.5 eV) with 1/3 of the energy being used as internal excitation of H$_2$, 1/3 being transferred to the grain and 1/3 being converted into kinetic energy. The fraction of H$_2$ formed in each ro-vibrational levels follows

\begin{equation}
f_{vJ}(T) = \frac{g_{J} \exp( -E_{vJ}/T)}{\sum g_{J} \exp( -E_{vJ}/T)}
\end{equation}

where $T=8734$K.  As noted in \citet{LePetit06}, this is different than assuming a Boltzmann distribution at $T=17322$ K (i.e. $\sim 1.5$eV), which would be equivalent to assume that approximately half (i.e. $\sim 2.5$ eV) of the  energy released at formation is used as internal excitation of H$_2$. For the ortho/para conversion on grains, we follow the approach of \citet{LeBourlot00} and consider that conversion occurs with an efficiency set by 

\begin{equation}
\eta_c = \exp\big( -\tau_c k_\text{ev})
\end{equation}

where $\tau_c$ is the conversion timescale and $k_\text{ev}$ the evaporation rate of H$_2$ from the grains. The o/p conversion rate is then calculated by

\begin{equation}
k_\text{o/p} = \alpha_{\text{H}_2} v_{\text{H}_2} \langle \sigma_d n_d \rangle \eta_c
\end{equation}

where $\alpha_{\text{H}_2}$ is the H$_2$ sticking probability taken from \citet{Matar10} and \citet{Chaabouni12}, $v_{\text{H}_2}$ is the thermal speed of H$_2$ molecules and $ \langle \sigma_d n_d \rangle$ the product of the dust cross sectional area and the dust density averaged over the grain size distribution.

%\subsection{Ortho/para conversion on grains}

\bibliography{bibliography}{}
\bibliographystyle{aasjournal}

\end{document}